\documentclass[showpacs,preprintnumbers,amsmath,twocolumn]{revtex4}

\usepackage{epsfig}
\usepackage{graphicx}
\usepackage{bm}
\usepackage{amsmath}
\usepackage{amssymb}

\renewcommand{\>}{\rangle}
\newcommand{\<}{\langle}
\newcommand{\ro}{\rho}

\newcommand{\rom}{\rho_{0\mu}}
\newcommand{\roam}{{\rho_{\alpha\mu}}}
\newcommand{\Mro}{{\hat{\rho}}}

\newcommand{\rr}{{\mathbf{r}}}
\newcommand{\rp}{{\mathbf{r}'}}

\newcommand{\Hk}{{\mathcal{H}}}
\newcommand{\F}{{\mathcal{F}}}
\newcommand{\sa}{{{\hat{\sigma}}_\alpha}}
\newcommand{\tm}{{{\hat{\tau}}_\mu}}
\newcommand{\ngg}{{{\hat{\nabla}}_\gamma}}

\newcommand{\pg}{{{{\partial}}_\gamma}}

\begin{document}


\title{Isospin coupling in time-dependent-mean-field theories and decay of isovector excitations}

\author{C. Simenel$^{1,2,3}$, Ph. Chomaz$^{2}$, T. Duguet$^{3,4}$}

\affiliation{$^{1}$ DSM/DAPNIA, CEA SACLAY, F-91191
Gif-sur-Yvette, France} \affiliation{$^{2}$ GANIL, B.P. 5027,
F-14076 CAEN Cedex 5, France} \affiliation{$^{3}$ NSCL, MSU, East
Lansing, Michigan 48824, USA} \affiliation{$^{4}$ Physics and
Astronomy Department, MSU, East Lansing, Michigan 48824, USA}

\date{\today}

\begin{abstract}

We show that isospin non-diagonal terms should appear in the mean
field Hamiltonian when neutron-proton symmetry is broken. They
give rise to charge mixing in the single-particle wave-functions.
We study the Time Dependent Hartree-Fock response of a
charge-exchange excitation which generates a charge mixing in Ca
isotopes. We find an enhancement of the low energy proton emission
 in neutron-rich isotopes interpreted in terms of a charge
 oscillation below the barrier.

\end{abstract}

\maketitle

Atomic nuclei are a good laboratory to test quantum mechanics of
few-body systems. The nucleon-nucleon ($NN$) interaction can be
interpreted in terms of meson exchange and depends on the quantum
numbers of the nucleons (spin, isospin, parity...). For instance
two protons or two neutrons cannot exchange a single charged pion,
unlike a proton-neutron pair. The $NN$ force can be divided into a
central, a spin-orbit and a tensor part \cite{oku58}, the latter
being mainly mediated by pions. The description of the structure
and the dynamic of nuclei through mean-field-like methods
\cite{bender03a} makes use of effective interactions such as the
Gogny \cite{gog80} or the Skyrme \cite{sky56} forces, taking care
of both the renormalization of the $NN$ interaction in the medium
and of the three-body force. In such models, the residual tensor
component is rarely explicitly included but the renormalization of
the central and spin-orbit parts by the original tensor, and in
particular the saturating effect of the latter, is incorporated on
a phenomenological basis through the density dependence and the
fitting procedure of the force. This is an important aspect since
the tensor force contributes strongly to the binding and
spin-orbit splittings of nuclei~\cite{pieper01a}.

However, treated fully or partially, the exchange of charged pions
can only be properly accounted for in a mean-field-like theory
through the breaking of isospin symmetry of single-particle (s.p.)
states. Indeed, including important parts of the residual
interaction through breaking of symmetries is a key concept of
self-consistent mean-field methods. However, mixing the isospin
projection of s.p. states leads to a many-body state with no good
charge any more. Such a broken symmetry is to be restored in
structure calculations since the nucleus is meant to have isospin
projection (i.e. the charge) as a good quantum number. Sugimoto
{\it et al} \cite{sug04} showed recently that a Hartree-Fock
(HF)~\cite{har28,foc30} calculation allowing for the explicit
mixing of protons and neutrons (as well as of parity), and
followed by the full Variation After Projection technique
\cite{pei57}, was indeed a powerful tool to include correlations
associated with the explicit treatment of the tensor force. Such
involved calculations can only be applied to very light nuclei
however. Also, the complete Hartree-Fock-Bogoliubov (HFB)
formalism including charge mixing in both the particle-hole (p-h)
and the particle-particle (p-p) channels was derived recently by
Perli\`nska {\it et al}~\cite{per04} within the Density Functional
Theory framework. In this context, both the neutron-proton pairing
and the correlations associated with the tensor force in the p-h
channel could be studied. However, no calculation has been
performed so far.

Symmetry breaking can also affect reaction mechanisms. When
correctly introduced and restored in the initial state, it may
account for the propagation of ground-state correlations.
Moreover, reactions may induce an explicit symmetry breaking. For
instance, the wave function of one of the collision partners in a
Charge-Exchange (CE) reaction becomes a superposition of isobaric
nuclei. While the wave function describing both collision partners
has to conserve charge exactly, both nuclei are entangled after
the collision, i.e. the wave function obtained by projection on
one of the two separated products have not a good isospin before
any detection takes place. Each component of the isospin $T_3$
affects the evolution of the others through the non linearities of
the mean field.

Mean-field methods should be generalized to take into account such
dynamical symmetry breaking induced by the reaction. The aim of
the present letter is to include, for the first time,
isospin-symmetry breaking in Time-Dependent Hartree-Fock (TDHF)
calculations \cite{dir30} within the Skyrme energy functional
framework. As a first application, the effect of correlations
associated with s.p. isospin mixing on the nucleon emission
following a CE excitation are studied.

In mean-field methods, allowing for s.p. isospin mixing translates
into the appearance of non-diagonal terms in the HF Hamiltonian in
isospin space. We derive such terms starting from an effective
Skyrme interaction \cite{bender03a}:

\begin{eqnarray}
{\hat{v}} &=&t_0(1+x_0{\hat{P}}_\sigma) \, {\hat{\delta}}
+\frac{t_1}{2}(1+x_1{\hat{P}}_\sigma)\left({\hat{\delta}}\,\,
{\mathbf{{\hat{k}}}}^2+{\mathbf{{\hat{k}}'}}{}^{2}\,\,{\hat{\delta}}
\right)\nonumber \\
& +&t_2(1+x_2{\hat{P}}_\sigma) \, {\mathbf{{\hat{k}}'}} . \,
{\hat{\delta}} \,\,{\mathbf{{\hat{k}}}}
+\frac{t_3}{6}(1+x_3{\hat{P}}_\sigma)
\,\,\rho^\beta ({\mathbf{{\hat{R}}_{12}}})\,\,{\hat{\delta}}\nonumber \\
& +& iV_{so}\left({\mathbf{{\hat{\sigma}}_1}}+
{\mathbf{{\hat{\sigma}}_2}}\right) .
 \, {\mathbf{{\hat{k}}'}}\times{\hat{\delta}}\,\, {\mathbf{{\hat{k}}}}
\, \, \, , \label{Eq_skyrme}
\end{eqnarray}

where ${\hat{\delta}}=\delta({\mathbf{{\hat{r}}_1}}
-{\mathbf{{\hat{r}}_2}})$,
${\mathbf{{\hat{R}}_{12}}}=({\mathbf{{\hat{r}}_1}}+{\mathbf{{\hat{r}}_2}})/2$,
${\mathbf{\hat{k}}}=\frac{{\mathbf{\hat{\nabla}}_1}-{\mathbf{\hat{\nabla}}_2}}{2i}$
acts on the right, while its Hermitian conjugated
${\mathbf{\hat{k}'}}$ acts on the left and
${\hat{P}}_\sigma=\frac{1}{2}(1+{\mathbf{{\hat{\sigma}}_1}}.{\mathbf{{\hat{\sigma}}_2}})$
is the spin exchange operator. In Eq. (\ref{Eq_skyrme})
$\rho({\mathbf{\hat{r}}})$ is the total local density defined from
the one-body density matrix $\<\rr sq|{\hat{\ro}}|\rp
s'q'\>=\sum_h \, \varphi_h^*(\rp s'q') \, \varphi_h(\rr sq)$,
where $ \varphi_h(\rr sq) = \<{\mathbf{r}}sq|h\>$ denotes the
component with a spin $s$ and isospin $q$ of the occupied s.p.
wave-function $h$. It is convenient to decompose ${\hat{\ro}}$
into its spin-isospin components acting only in coordinate space:
\begin{equation}
{\hat{\rho}}_{\alpha\mu}=\sum_{ss'qq'} \<sq|{\hat{\rho}}|s'q'\>
\<s|{\hat{\sigma}}_\alpha|s'\>\<q|{\hat{\tau}}_\mu|q'\> \, \, \, ,
\label{Eq_spin-isospin}
\end{equation}
where the $\alpha$ index relates to the spin degree of freedom and
$\mu$ to the isospin one ($\gamma$ or $\eta$ are kept for the
3D-space). Whenever $\alpha$ ($\mu$) is $0$, only the scalar
(isoscalar) part of the density is considered, i.e.
${\hat{\sigma}}_\alpha$ (${\hat{\tau}}_\mu$) is replaced by the
identity, ${\hat{\sigma}}_0=1_s$ (${\hat{\tau}}_0=1_q$). In Eq.
(\ref{Eq_spin-isospin}), densities with $\mu=1,2$ are non-diagonal
in isospin and will then describe the charge mixing. The HF
ground-state energy associated with the Skyrme force defined in
Eq.~(\ref{Eq_skyrme}) can be written as an integral of a local
energy density ${\mathcal{H}}({\mathbf{r}})$ which is a functional
of the spin-isospin local densities
$\rho_{\alpha\mu}({\mathbf{r}}) =
\<\rr|{\hat{\rho}}_{\alpha\mu}|\rr\>$, kinetic densities
$T_{\alpha\mu}(\rr)=
\left[\sum_\gamma\nabla_\gamma{\nabla'}_\gamma
\<\rr|{\hat{\rho}}_{\alpha\mu}|\rp\>\right]_{\rr=\rp}$ and current
densities $J_{\gamma\alpha\mu}(\rr) = \frac{1}{2i}\left[\left(
\nabla_\gamma-{\nabla'}_\gamma\right)\<\rr|{\hat{\rho}}_{\alpha\mu}|\rp
\>\right]_{\rr=\rp}$. In order to shorten the equations, the $\rr$
dependence is not written explicitely in the following. The
contributions to ${\mathcal{H}}$ associated with $t_0$, $t_1$,
$t_2$ and $t_3$ terms in Eq. (\ref{Eq_skyrme}) have similar
structures:
\begin{equation}
\Hk_k = C_k \sum_{\alpha,\mu=0}^3
d_{k}^{\alpha\mu}\F_{k}^{\alpha\mu} \, \, \, , \nonumber
\end{equation}
with  $ d_{2}^{00}\!=\!5\!+\!4x_2$, $d_{2}^{\alpha 0}|_{\alpha\neq
0}\!= \!d_{2}^{0\mu}|_{\mu\neq 0}\!=\!2x_2\!+\!1$,
$d_{2}^{\alpha\mu}|_{\alpha\mu\neq 0}\!=\!1$ and, for $k\in
\{0,1,3\}$, $d_{k}^{00}\!=\!3$, $ d_{k}^{\alpha 0}|_{\alpha\neq
0}\!=\!2x_k\!-\!1$, $ d_{k}^{0\mu}|_{\mu\neq 0}\!=\!-2x_k\!-\!1$
and  $d_{k}^{\alpha\mu}|_{\alpha\mu\neq 0}\!=\!-1$. The functions
$C_k(\rr)$ are defined by $C_0\!=\!\frac{t_0}{8}$,
$C_1\!=\!\frac{-t_1}{64}$, $C_2\!=\!\frac{t_2}{64}$,
$C_3\!=\!\frac{t_3}{48}\ro_{00}^{\beta}$ and the functionals $\F$
by $\F_0^{\alpha\mu}\!=\!\F_3^{\alpha\mu}\!=\!\rho^2_{\alpha
\mu}$,
\begin{equation}
\F_1^{\alpha\mu}\!=\!3\roam \Delta \roam
\!-\!4 \roam T_{\alpha\mu}\!+\!4{J^2_{\gamma {\alpha\mu}}} \, \, ,
\nonumber
\end{equation}
and $\F_2^{\alpha\mu}\!=\! 4 \roam \Delta \roam \!-\!
\F_1^{\alpha\mu}$. The spin-orbit term is:
\begin{equation}\Hk_{so}[\Mro] = \frac{-V_{so}}{4} \sum_{\mu=0}^3 d_{so}^\mu \F_{so}^{\mu} \, \, \, ,
\end{equation}
with $d_{so}^\mu\!=\!1\!+\!2\delta_{\mu0}$ and
\begin{equation}
\F_{so}^\mu\!=\! \sum_{\gamma\eta\alpha=1}^3
\varepsilon_{\gamma\eta\alpha} \left(J_{\gamma 0\mu}\nabla_{\eta}
\roam \!+\! J_{\gamma \alpha\mu}\nabla_{\eta} \ro_{0\mu} \right)
\, \, \, , \nonumber
\end{equation}
where $\varepsilon_{\gamma\eta\alpha}$ is the antisymmetric
tensor.

The HF Hamiltonian is obtained through the functional derivative
${\hat{h}}[\Mro]=\partial E[\Mro]/ \partial \Mro^T$. One expresses
the variation of the energy through the variation of the various
local densities defined above. One can show in the same way as in
Ref. \cite{eng75} that their contributions to the HF Hamiltonian
are $\delta \roam \rightarrow {\hat{\sigma}}_\alpha
{\hat{\tau}}_\mu$, $f\delta T_{\alpha\mu} \rightarrow -
{\hat{\sigma}}_\alpha {\hat{\tau}}_\mu \sum_\gamma
{\hat{\nabla}}_\gamma f{\hat{\nabla}}_{\gamma}$ and $f \delta
J_{\gamma\alpha\mu} \rightarrow \frac{1}{2i} {\hat{\sigma}}_\alpha
{\hat{\tau}}_\mu \left( {\hat{\nabla}}_\gamma f + f
{\hat{\nabla}}_\gamma\right)$ where the ${\hat{\nabla}}$ operators
act on each term sitting on their right, including the wave
functions. Finally the HF Hamiltonian reads as:

\begin{eqnarray}
{\hat{h}} & = & -\frac{\hbar^2}{2m} {\hat{\Delta}}+
\frac{\beta}{\ro} \Hk_3 + 2\sum_{ k,\alpha,\mu=0}^3 C_k
d_{k}^{\alpha\mu} \sa\tm {\hat{G}}_k^{\alpha\mu}  \nonumber \\
& -&  \frac{V_{so}}{4} \sum_{\alpha=1, \mu=0 }^3 d_{so}^\mu \tm
({\hat{G}}_{so}^{\alpha\mu}+\sa {\hat{G}}_{so}^{ 0 \mu})  \, \, \,
, \label{Eq_fullh}
\end{eqnarray}

with the operators ${\hat{G}}$ given by ${\hat{G}}_0^{\alpha\mu}=
{\hat{G}}_3^{\alpha\mu}=\roam$, ${\hat{G}}_2^{\alpha\mu}=4\Delta
\roam-{\hat{G}}_1^{\alpha\mu}$ and

\begin{eqnarray}
{\hat{G}}_1^{\alpha\mu} &=& 3\Delta \roam-2 T_{\alpha\mu} \nonumber \\
& + &  \sum_\gamma 2\ngg \roam \ngg -2i (\ngg J_{\gamma\alpha\mu}+
J_{\gamma\alpha\mu } \ngg) \nonumber \\
{\hat{G}}_{so}^{\alpha\mu}&=& \sum_{\gamma\eta}
\varepsilon_{\gamma\eta\alpha}
\left(\frac{1}{2i}(\ngg{{\partial}}_\eta \roam+{{\partial}}_\eta
\roam\ngg)-{{\partial}}_\eta J_{\gamma\alpha\mu}\right) \nonumber
\, \, \, ,
\end{eqnarray}
where the derivative ${\pg}$ and $\Delta$ acts only on the
function next to them. Everything derived above is valid both for
HF and TDHF calculations which corresponds to
$\left[{\hat{h}}[{\hat{\rho}}(t)],{\hat{{{\rho}}}}(t)\right] =
i\hbar{\partial_t}{\hat{{{\rho}}}}(t)$~\cite{eng75}.

\begin{figure}
\begin{center}
\epsfig{figure=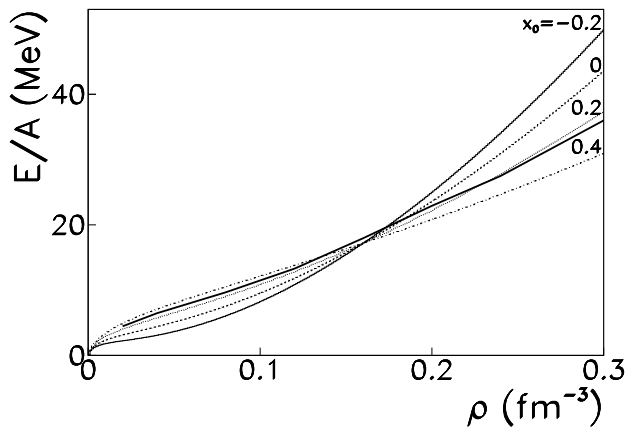,width=9cm} \caption{EoS of pure neutron
matter. Variational calculation with Eq.
(\ref{Eq_skyrme_simplified}) with different values of $x_0$.}
\label{EOS}
\end{center}
\end{figure}

The breaking of the isospin symmetry leads to a non diagonal
$\hat{h}$ in isospin space because of the ${\hat{\tau}}_{1,2}$
operators in Eq.(\ref{Eq_fullh}). Through the dynamics, such a
non-diagonal Hamiltonian will lead to an isospin oscillation
analogous to the Larmor precession of spins in an unaligned
magnetic field or to particle oscillation in a non diagonal mass
operator. In order to quantitatively illustrate the phenomenon of
isospin mixing in the case of a CE excitation, let us use a
simplified version of the complete Skyrme force introduced in
Eq.~(\ref{Eq_skyrme}):

\begin{equation}
{\hat{v}} =
\left(f[\ro_{00}]+f_\sigma[\ro_{00}]{\hat{P}}_\sigma\right)\hat{\delta}
\, \, \, , \label{Eq_skyrme_simplified}
\end{equation}

with $f[\ro_{00}] = t_0 + \frac{t_3}{6} \rho_{00}^{\beta}$ and
$f_\sigma[\ro_{00}] = t_0 x_{0} + \frac{t_3}{6} x_{3}
\rho_{00}^{\beta}$. The parameters $t_0=-1803.38$ MeV.fm$^{3}$ and
$t_3=12912.80$ MeV.fm$^{3(1+\beta)}$ have been adjusted to
reproduce the empirical saturation point of infinite symmetric
matter whereas $\beta = 1/3$ was chosen to have a reasonable
compressibility $K_{\infty} = 236.2 \,$MeV. As momentum-dependent
and spin-orbit terms are dropped, the nucleon effective mass
$m^{\ast}$ is equal to the bare one and the scalar time odd
current density is not generated by the force, while spin-orbit
splittings and some magic numbers are not reproduced in finite
nuclei. Asking for a value of the symmetry energy $a_{s} = 32.5$
MeV at saturation density $\rho_{sat}=0.16$ fm$^{-3}$ gives the
relation $x_3 = 1.544 \,x_0 + 0.161$. Different values of $x_{0}$
were tried in order to match the Equation of State (EoS) of
infinite neutron matter predicted by Akmal {\it et al.} through
variational chain summation methods~\cite{akmal} (thick-solid line
on Fig.~\ref{EOS}). The value of $x_0 = 0.2 $ for which the EoS of
neutron matter obtained from microscopic calculations is well
reproduced was finally used (see Fig. \ref{EOS}). In view of the
commonly accepted value chosen the symmetry energie and the fair
reproduction of the EoS of neutron matter, one can expect the
isovector properties of the simplified Skyrme force and the
isospin mixing arising from it to be reasonable. In addition, the
proton and neutron separation energies are close to the
experimental values for the $^{40}$Ca: $S_p=8.25$ MeV and
$S_n=16.56$ MeV. However, going far from stability the separation
energies are not as good. For $^{60}$Ca, a HFB-Lipkin-Nogami
calculation gives $S_p=24.33$ MeV and $S_n=3.57$ MeV \cite{dug05}
whereas we get $S_p=19.35$ MeV and $S_n=8.38$ MeV. The full Skyrme
force of Eq. \ref{Eq_skyrme} is needed to correct for this
discrepancy.

\begin{figure}
\begin{center}
\epsfig{figure=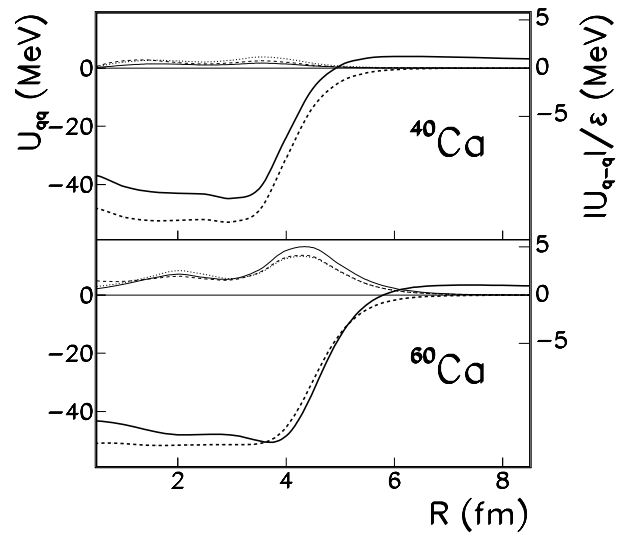,width=8cm} \caption{Proton (thick
solidline) and neutron (thick dashed line) diagonal potentials
$U_{qq}$ and isospin coupling modulous $|U_{q-q}|$ divided by the
boost strength $\varepsilon$ at 30, 360 and 690 fm/c (thin solid,
dashed and dotted lines respectively) for non spin-flip CE
excitations in $^{40}$Ca (up) and $^{60}$Ca (down). }
\label{fig:pot}
\end{center}
\end{figure}

We now write the TDHF potential associated with the force in
Eq.~(\ref{Eq_skyrme_simplified}), focusing on even-even nuclei. In
their ground state, only time-reversal invariant local densities
are non-zero. Thus, s.p. states with a good spin can be used. This
remains true during the dynamical evolution as long as we do not
break explicitly the spin symmetry in the wave function. With this
hypothesis, all terms with a $\alpha$ index vanish in Eq.
(\ref{Eq_fullh}). Finally, we are left with the potential matrix
elements:

\begin{eqnarray}
U_{qq}&=& \frac{3}{8}\frac{\partial}{\partial \ro_{00}}\left(f\ro_{00}^2\right)
-\frac{1}{8}\left(4q\ro_{03}+\sum_{\mu=1}^3 \rom^2\frac{\partial}{\partial \ro_{00}}\right) \nonumber \\
& & \left(f+2f_\sigma\right)+ \delta_{q,\frac{1}{2}} U_{c} \nonumber \\
U_{q-q}&=&-\frac{1}{4} \left( f+2f_\sigma \right) \eta_q \, \, \,
, \label{eq:pot_diag_undiag}
\end{eqnarray}
where $U_c $ is the Coulomb potential. The non-diagonal part
$U_{q-q}$, or  ``isospin coupling'' is proportional to the isospin
mixing density, $\eta_q=\frac{1}{2}\rho_{01}+iq\rho_{02} = \sum_s
\<\rr sq|{\hat{\ro}}|\rr s -\!q\>$. We can now study a non
spin-flip CE excitation generated by a collective isovector boost
applied at $t=0$ to an isospin-diagonal HF ground state
$|\phi_0\>$: $|\phi(0)\> = e^{-i\varepsilon {\hat{\bf \tau}}_{\bf
u}}|\phi_0\>$ where $\varepsilon$ and ${\bf u}$ denote the boost
strength and direction in isospin space, respectively. When ${\bf
u}$ is not aligned with the ${\hat{\tau}}_3$ axis, a CE takes
place and $|\phi(0)\>$ is then an antisymmetrized product of s.p.
wave-functions which mixed protons and neutrons. For $N=Z$ nuclei
the CE produced is quenched by the Pauli principle (and vanishes
if Coulomb is neglected). As explained earlier, the wave function
after the CE excitation is a linear combination of isobaric analog
nuclei. The influence on the dynamics of such a quantum
entanglement in isospin space is incorporated in the TDHF method
developped in the present letter.

\begin{figure}
\begin{center}
\epsfig{figure=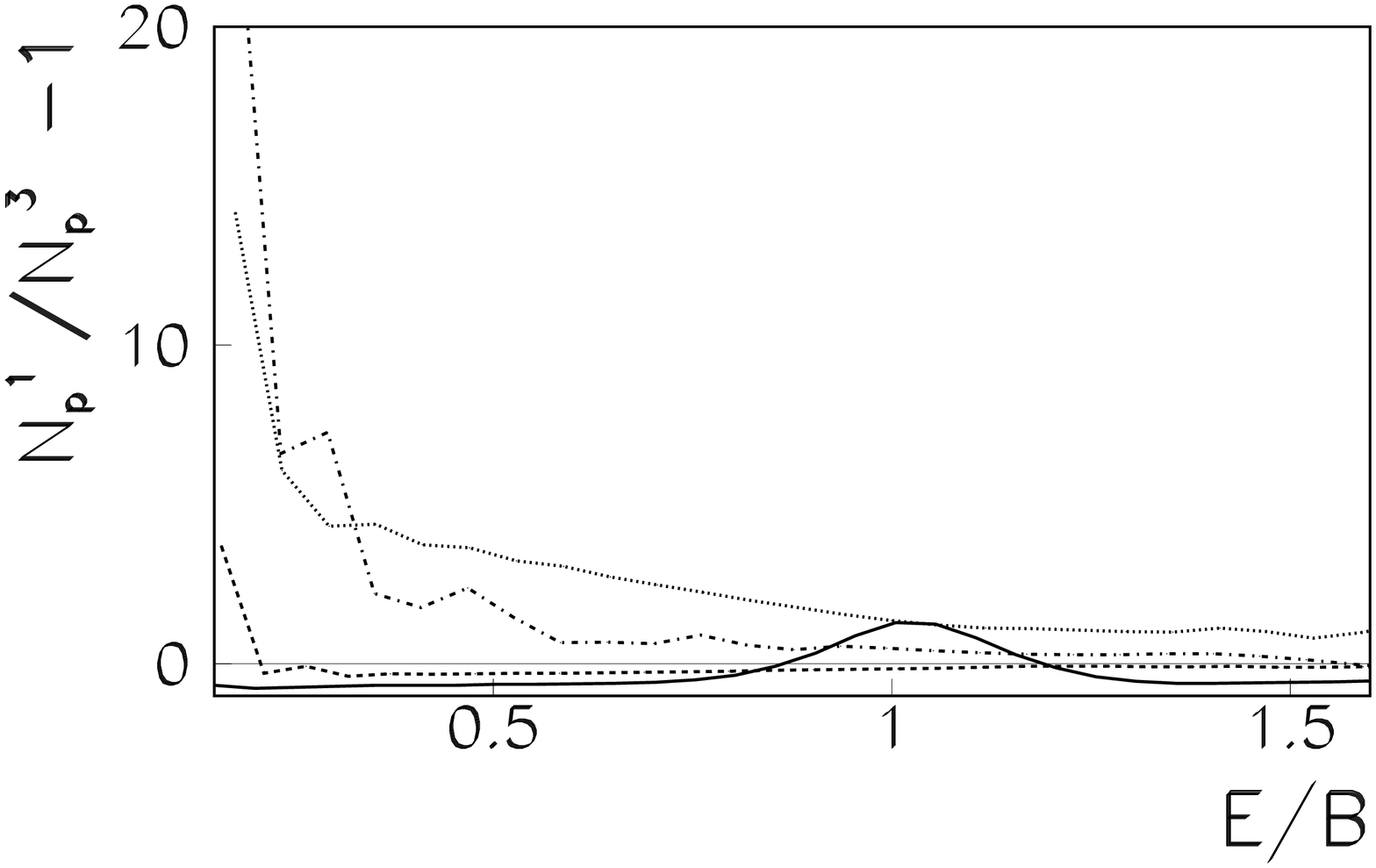,width=8cm} \caption{Ratio of the numbers
of emitted proton following an isovector excitation by the
${\tau_1}$ and $\tau_3$ operators as function of the proton energy
E divided by the barrier B for the $^{40-48-54-60}$Ca (solid,
dashed, dotted, dotted-dahed lines respectively).}
\label{fig:trans}
\end{center}
\end{figure}

We have computed numerically the TDHF evolution after a boost an
isovector applied on the ground state of $^{40}$Ca and $^{60}$Ca.
The simplified Skyrme  mean field  Eq. (\ref{eq:pot_diag_undiag})
is considered. If $\varepsilon << 1$, the diagonal potentials
($U_{qq}$) are almost constant in time while the isospin coupling
is linear in $\varepsilon$. The resulting potentials at different
times are plotted in Fig. \ref{fig:pot}. The small variation in
time of $|U_{q-q}|$ comes from the difference in energy of the
excited modes. $|U_{q-q}|$ is more intense in $^{60}$Ca than in
$^{40}$Ca as expected from the quenching of CE processes in the
N$=$Z nuclei. The isospin coupling is also more peaked at the
surface in $^{60}$Ca because of the isospin mixing in the
2$p$-1$f$ shells and decreases exponentially at the barrier.

Let us now present a possible effect of the isospin coupling on
the low energy proton emission. The particle emission looked at in
the present case originates from the entangled state prior to any
detection of the reaction partner ($\Leftrightarrow$ projection on
good isospin). An emitted proton (neutron) can come from an
incident proton (neutron) or an incident neutron (proton) which
has exchanged its isospin via the isospin coupling. At low energy,
where the proton emission is hindered by the Coulomb barrier, one
can expect an enhancement of the proton emission from neutron
which exchange their isospin at the surface. This is possible
because the isospin coupling does not fully vanish at the surface
(see Fig. \ref{fig:pot}). The excitation studied above does not
usually populate states above the nucleon threshold. To reach such
high excitation energy we applied on the HF ground state an
isovector-monopole boost $e^{-i\varepsilon {\hat{r}}^2{\hat{\bf
\tau}}_{\bf u}}$. Due to the $ {\hat{r}}^2$ operator, an important
part of the strength goes to the 2$\hbar \omega$ non spin-flip
Isovector Giant Monopole Resonance. Let us compare the numbers of
protons emitted at an energy E after a boost generated by $\tau_3$
(noted $N_p^{3}$) and by $\tau_1$ ($N_p^{1}$).
Fig.~\ref{fig:trans} shows the evolution of $N_p^{1}/N_p^{3}-1$
obtained for $^{40-48-54-60}$Ca. We choose $\varepsilon =0.01$
fm$^{-2}$ which is small enough to be in the linear regime.

For a $\tau_3$-boost, there is no isospin coupling whereas a
$\tau_1$ excitation produces an isospin coupling slightly more
peaked at the surface than in Fig.~\ref{fig:pot} due to the $r^2$
term in the boost. In this case the expectation value of the
charge is changed by $\Delta \<Z\>=-0.0044$ in $^{40}$Ca and 0.064
in $^{60}$Ca. Fig.~\ref{fig:trans} shows a strong enhancement of
the proton emission at energies well below the barrier except for
the $^{40}$Ca for which the isospin coupling remains small. Around
the barrier this effect is hindered by the difference in the
excitation energy spectra associated with the inelastic and the
charge exchange excitations. This enhancement of the low energy
proton is expected to be a general consequence of isospin
coupling. In fact we expect this effect to be strong in all nuclei
with large proton/neutron asymmetry.

To conclude, we have shown that isospin coupling appears in the HF
mean field when proton-neutron symmetry is broken. Its expression
was derived for a full Skyrme-like force. We used a simplified
Skyrme force adjusted on the isospin properties of nuclear matter
to get an approximation of the isospin mixing. As a first
application we performed TDHF calculations using this force to
study the effect of charge-exchange excitations on nucleon
emission. We found an enhancement of the proton emission below the
barrier for the neutron rich Ca interpreted in terms of nucleon
charge oscillation. More quantitative predictions are needed with
a 3D-TDHF code with a full Skyrme force to study complete
reactions, i.e. not only the evolution of one nucleus wave
function. The inclusion of the spin degree of freedom would also
allow studies of Gamow-Teller transitions and isospin and
spin-flip giant resonances. For the latter case however and for
any nuclear structure study in general, it is necessary to keep
the good value of the charge in average during the time evolution
and to project on it after.

We thank Y. Blumenfeld, J.A. Scarpacci, P. Bonche and M. Bender,
for useful discussions during this work.


\end{document}